\documentclass[prb,twocolumn,superscriptaddress,floatfix,showpacs,amsmath,amssymb]{revtex4}

\usepackage{multirow}
\usepackage{graphicx}
\usepackage{bm}
\usepackage{picture}
\usepackage{amsmath}
\usepackage{times}
\usepackage{color}
\usepackage{xcolor}
\usepackage{amsmath}
\usepackage{amssymb}
\usepackage{booktabs}
\usepackage{tabularx}
\usepackage{array}
\usepackage{gensymb}

\newcommand{\nto}{NiTiO$_3$}
\newcommand{\mto}{MnTiO$_3$}
\newcommand{\cto}{CoTiO$_3$}
\newcommand{\fto}{FeTiO$_3$}
\newcommand{\tn}{$T_{\rm N}$}

\newcommand{\mb}{$\mu_{\rm B}$}

\newcommand{\etal}{\textit{et~al.}}

\begin{document}

\title{Magneto-structural coupling in ilmenite-type \nto: a combined diffraction and dilatometry study }

 \author{K.~Dey}
\email[email:]{kaustav.dey@kip.uni-heidelberg.de}
\affiliation{Kirchhoff Institute of Physics, Heidelberg University, INF 227, 69120, Heidelberg, Germany}

\author{S.~Sauerland}
\affiliation{Kirchhoff Institute of Physics, Heidelberg University, INF 227, 69120, Heidelberg, Germany}

\author{B.~Ouladdiaf}
\affiliation{Institut Laue-Langevin, CS20156, 38042 GRENOBLE Cedex 9 - France}	
	
\author{K.~Beauvois}
\affiliation{Institut Laue-Langevin, CS20156, 38042 GRENOBLE Cedex 9 - France}

\author{H.~Wadepohl}
\affiliation{Institute of Inorganic chemistry, Heidelberg University, 69120, Heidelberg, Germany}

\author{R.~Klingeler}
\affiliation{Kirchhoff Institute of Physics, Heidelberg University, INF 227, 69120, Heidelberg, Germany}\affiliation{Center for Advance Materials (CAM), Heidelberg University, INF 227, D-69120, Heidelberg, Germany}

	\date{\today}
\begin{abstract}
			{
We report the ground state magnetic structure and in-field magnetostrictive effects of \nto\ studied by means of zero field and in-field single crystal neutron diffraction, magnetization and high-resolution dilatometry experiments. Zero-field neutron diffraction on \nto\ single crystals reveal an easy-plane antiferromagnet with a multidomain ground state. Upon application of external magnetic fields, neutron diffraction shows the evolution of domains with spins perpendicular to the applied field. The rotation of spins in the multidomain state exhibits pronounced lattice changes in the magnetostriction measurements. We see magnetization and magnetostriction measurements sale with each other in the multidomain state revealing the strong coupling of spins to the lattice. 
			}
	\end{abstract}
		
\maketitle
\section{Introduction}

Layered honeycomb magnets have been a great avenue for exciting and rich physics since time immemorial. The recent theoretical and experimental studies into Kitaev quantum spin-liquid in Co-based honeycomb materials\cite{Cava_2020,Khaliullin_2018}, Dirac magnons\cite{Diracmagnon_2018} and topological spin excitations\cite{Topologicalexci_2018} in honeycomb ferromagnets, non-reciprocal magnons in honeycomb antiferromagnets\cite{Nonreciprocalmagnons_2020}, zig-zag~\cite{Kurbakov_2017} and incommensurate~\cite{NAKUA_1995,koo_2017} spin ground states or 2D magnetism in Van-der-Waals materials\cite{VdW_2018} have resulted in enormous interest in these class of materials. Moreover, the spin-lattice coupling in several honeycomb magnets such as Fe${_4}$Nb${_2}$O${_9}$\cite{martin_2018}, Na$_3$Ni$_2$SbO$_6$~\cite{Johannes_2017}, and Co${_4}$Nb${_2}$O${_9}$\cite{Arima_2019} have resulted in significant magnetoelectric coupling and hence motivating possible technological applications.
 
Ilmenite titantes with chemical formula $M$TiO$_3$ ($M$ = Mn, Fe, Co, Ni) form an isostructural series of antiferromagnetic (AFM) compounds where magnetic $M^{2+}$ ions in the basal $ab$-plane exhibit a buckled honeycomb-like structure. The $M^{2+}$ ions are interconnected via oxygen ions (O$^{2-}$) leading to $M$-O-$M$ as the dominant superexchange pathway~\cite{Goodenough}. Along $c$ axis the crystal structure exhibits alternating layers of corner sharing TiO$_6$ and $M$O$_6$ octahedra resulting in relatively weaker $M$-O-Ti-O-$M$ superexchange pathways. Depending on the single ion-anisotropies of the respective metal ions, various magnetic ground states are realised in ilmenites, for example uniaxial AFM ground state with spins pointing along $c$-axis in \mto\cite{Shirane_1959} whereas an easy-plane type AFM with spins lying in the $ab$-plane for \nto\ and \cto\cite{Newnham} respectively. 
 
Although, these compounds have been rigorously investigated since 1950s~\cite{Shirane_1959,Stickler,Heller,Goodenough,Akimitsu_1970}, recent studies evidencing linear magnetoelectric coupling in \mto\cite{Mufti_2011}, large spontaneous magnetostriction in \fto\cite{charilou}, magnetodielectric and magnetoelastic coupling in \nto\cite{Harada,dey2020} and \cto\cite{DUBROVIN2020}, respectively, as well as the observance of Dirac magnons in \cto\cite{Yuan_2020,elliot_2020} have peaked enormous interest in these class of materials. 

The least investigated compound among the ilmenites family, i.e., \nto , develops long-range AFM order at \tn ~=~22.5~K~\cite{Stickler,Watanabe,Harada,dey2020}. Recent studies of the dielectric permittivity and the thermal expansion show a pronounced magnetodielectric effect~\cite{Harada} as well as distinct significant magnetoelastic coupling~\cite{dey2020}. Notably, at \tn , there is single energy scale dominantly driving the observed structural, magnetic and dielectric anomalies~\cite{dey2020}. In this report, we study in detail the magneto-structural coupling of \nto\ by means of single crystal X-ray and neutron diffraction and high-resolution dilatometry. We observe by means of single-crystal neutron diffraction that the macroscopic structural symmetry($R$-3) is retained down to the lowest measured temperature of 2~K within the experimental resolution. In addition, the magnetic ground state of \nto\ is solved. At \tn , in addition to long-range AFM order, a significant lattice distortion evolves revealing large spontaneous magnetostriction in \nto . In applied magnetic fields, the multi-domain ground state evolves to a spin-reoriented single domain state characterized by spins aligned perpendicular to the applied magnetic field. Magnetostriction measurements in the low-field region show pronounced effects due to magnetoelastic domains and remarkably scales with magnetization measurements confirming both significant magneto-structural coupling and the magneto-structural domain model in \nto. 

\section{Experimental methods}

Macroscopic single crystals of \nto\ have been grown by means of the optical floating-zone technique in a four mirror optical floating-zone furnace (CSC, Japan) equipped with 4$\times$150~W halogen lamps. Details of the growth process and characterization the single crystals have been published previously~\cite{dey2020}. Single crystal X-ray intensity data were obtained at 100~K with an Agilent Technologies Supernova-E CCD 4-circle diffractometer (Mo-K$\alpha$ radiation $\lambda$=0.71073~\AA, micro-focus X-ray tube, multilayer mirror optics). Static magnetisation $\chi=M/B$ was studied in magnetic fields up to 5~T in a Quantum Design MPMS-XL5 SQUID magnetometer. The relative length changes $dL_i/L_i$ were studied on a cuboid-shaped single crystal of dimensions $2 \times 1.85 \times 1~$mm$^{3}$ by means of a three-terminal high-resolution capacitance dilatometer~\cite{dil,Johannes_2017}. Magnetostriction, i.e., field-induced length changes $dL_i(B)/L_i$, was measured at several fixed temperatures in magnetic fields up to 15~T and the longitudinal magnetostriction coefficient $\lambda_i = 1/L_i\cdot dL_i(B)/dB$ was derived. The magnetic field was applied along the direction of the measured length changes.

Single crystal neutron diffraction experiments were performed up to 6~T magnetic fields on the D10 beamline of the Institut Laue-Langevin (ILL) at Grenoble, France. To determine the magnetic ground state at $B=0$~T, the four-circle configuration was used with a 96$\times$96 mm$^2$ two-dimensional microstrip detector. An incident wavelength of 2.36~\AA\ using a vertically focusing pyrolytic graphite (PG)(002) monochromator was employed. A pyrolytic graphite filter was used in order to suppress higher-order contamination to 10$^{-4}$ times that of the primary beam intensity. Measurements were made in the temperature range 2-50~K. The magnetic field-driven evolution of the magnetic structure at $T=2$~K was studied by mounting the sample in a 6~T vertical cryomagnet and aligned to within 1$^\circ$ of magnetic field. The magnetic field was applied along the $b$-axis limiting the scattering to the $(H,0,L)$ plane. 

\section{Experimental results}

\subsection{Single-crystal X-ray Diffraction}

To the best of our knowledge, the earlier studies of the ilmenite-type \nto\ crystal structure have been limited to powder diffraction experiments only~\cite{barth_1934,Shirane_1959,dey2020}. We have re-investigated the crystal structure by means of single-crystal high resolution XRD at 100~K, using Mo K$\alpha$ radiation ($\lambda$ = 0.71073~\AA). A single crystal splinter of size 0.16$\times$0.14$\times$0.01 mm$^3$ was broken of from larger specimen and used for data collection. A full shell of intensity data was collected up to 0.4~\AA\ resolution (24180 reflections, 1028 independent [$R_{int} = 0.05$] of which 1024 were observed [$I > 2\sigma(I)$]). Detector frames (typically $\omega$, occasionally $\phi$-scans, scan width 0.5$^\circ$) were integrated by profile fitting\cite{kabsch2001}. Data were corrected for air and detector absorption, Lorentz and polarization effects\cite{crysalis} and scaled essentially by application of appropriate spherical harmonic functions~\cite{crysalis,blessing1995,scale3}. Absorption by the crystal was treated numerically (Gaussian grid)~\cite{scale3,busing1957}. An illumination correction was performed as part of the numerical absorption correction\cite{scale3}. Space group $R-3$ was assigned based on systematic absences and intensity statistics (refined obverse centered unit cell on hexagonal axes, Hall group $-R 3$, $a = 5.02762(6), c = 13.76711(17) \AA$, $V = 301.369(8) \AA^3, Z = 6$). This choice was confirmed by analysis of the symmetry of the phases obtained ab initio in P1. The structure was solved by intrinsic phasing \cite{Sheldrick_2012,ruff_2014,sheldrick_2015} and refined by full-matrix least-squares methods based on $F^2$ against all unique reflections\cite{Sheldrick_2012a, Robinson_1988, sheldrick_2008, sheldrick_2015a}. Three somewhat different models were employed for the atomic structure factors $f_{at}$ within the ISA approximation: conventional $f_{at}$ calculated with neutral atoms\cite{Brown_2004} for Ni, Ti and O (model A) and two “ionic” models\cite{Brown_2004} ($f_{at}$ for  Ni$^{2+}$, Ti$^{4+}$ taken from ref.\onlinecite{Brown_2004} and O$^{2-}$ from ref.\onlinecite{Azavant_1993}  (model B) or ref.\onlinecite{Morel_2016}, respectively (model C)). An empirical secondary extinction correction \cite{Robinson_1988,Larson} was applied in each case but proved insignificant. The different models refined to essentially the same structure, with only insignificant differences in key parameters like atom coordinates, $R$ factors, $U_{eq}$ for all atoms and residual electron density. Ni-O and Ti-O bond lengths agreed within one standard deviation. There was no evidence of cation mixing and fully occupied sites were employed for all atoms. The results confirm the assignment of the space group and improve on the accuracy of the crystallographic parameters previously obtained from powder XRD and neutron data \cite{barth_1934, dey2020, Shirane_1959}. Fractional atomic coordinates, Wyckoff positions, site occupation and equivalent isotropic displacement parameters for model A are listed in table I \cite{ICSD}.

\begin{table}[h]
	\centering
	\caption{Fractional atomic coordinates, Wyckoff positions, site occupation and equivalent isotropic displacement parameters (\AA$^2$) for \nto\ at 100~K as obtained from refinement of model A. (Note: (1) These co-ordinates are correct but do not form uniquely bonded set; (2) $^a$ $U_{eq}$ is defined as one third of the trace of the orthogonalized $U_{ij}$ tensor. The anisotropic displacement factor exponent takes the form: $-2 \pi^2[h^2a^{*2}U_{11} + ... + 2hka^*b^*U_{12}]$.)} \vspace{1mm}
	\begin{tabular}{lcccccc}
	  \hline \hline
		Atom & Site & x & y & z & sof & U$_{eq}$$^a$ \\ 
		\hline
	    Ni &  6c   &      0         &  0	      &  0.35051(2)   &  1   &  0.00308(2) \\
		Ti &  6c   &      0         &  0	      &  0.14422(2)   &  1   &  0.00297(3) \\
	    O  &  18f  &  0.35198(8)    &  0.03455(8) &  0.08662(2)   &  1   &  0.00421(4) \\
      \hline \hline 
   	\end{tabular}
	\label{tab_SXRD}
\end{table}

\subsection{Single-crystal neutron diffraction}

The crystal structure at lower temperatures and the magnetic ground state of \nto\ were determined by means of single-crystal neutron diffraction. At 50~K, 110 nuclear Bragg reflections were collected. Appropriate correction for extinction, absorption, and Lorentz factor was applied to all the nuclear Bragg peaks. All the nuclear peaks at 50~K were successfully indexed in the $R$-3 space group with lattice parameters $a$ = 5.03~\AA\  and $c$ = 13.789~\AA.

In order to clarify the magnetic structure, preliminary reciprocal-space scans (not shown here) were performed at 2~K along the $(0,0,L), (H,0,0)$, and $(H,K,0)$ directions. The scans reveal a peak of significant intensity emerging at (0,0,1.5), indicative of the propagation vector \textbf{k} = (0,0,1.5). In order to determine the detailed magnetic structure, integrated intensities of 187 nuclear reflections allowed within the space group $R$-3 and 292 satellite magnetic reflections were collected at 2~K. The nuclear structure was firstly refined using FULLPROF program within the $R$-3 space group. The results of refinement are listed in table~\ref{tab_nuc2K} and the observed and calculated intensities from the Rietveld fits are shown in Fig.~\ref{neutrostruct}(a). No peak splitting or significant broadening was observed within the experimental resolution in respective 2~K nuclear reflections as compared to 50~K, indicating that the macroscopic $R$-3 symmetry is maintained until the lowest measured temperatures. The nuclear Bragg peaks show no temperature dependence between 2~K and 50~K excluding \textbf{k} = (0,0,0).

\begin{table}[ht]
	\centering
	\caption{Parameters for the nuclear structure of \nto\ measured at 2~K obtained from refinements of single-crystal neutron diffraction data. The isotropic temperature factors ($B$) of all atoms were refined. [Space group: $R$-3 (148); Lattice parameters: $a$ = $b$ = 5.0229(1)~\AA, $c$ = 13.7720(1)~\AA, $\alpha$ = $\beta$ = 90$^{\circ}$, $\gamma$ = 120$^{\circ}$.}\vspace{1mm}
	\begin{tabularx}{1\linewidth} {  >{\raggedright\arraybackslash}X  >{\centering\arraybackslash}X  >{\centering\arraybackslash}X  >{\centering\arraybackslash}X >{\centering\arraybackslash}X >{\centering\arraybackslash}X  }
	  \hline \hline
		Atom & Site & x & y & z  & $B_{iso}$(\AA$^2$) \\ 
    	\hline 
	    Ni1  &  6c    &      0         &  0	          &  0.3537(2)      &  0.00748 \\
	    Ti1  &  6c    &      0         &  0	          &  0.1338(5)      &  0.06643 \\
		O1   &  18f   &  0.3344(6)     &  0.0052(1)   &  0.2466(2)      &  0.09830 \\
	   \hline \hline
	\end{tabularx}
	\label{tab_nuc2K}
\end{table}

\begin{figure}[ht]
\centering
\includegraphics [width=1\columnwidth] {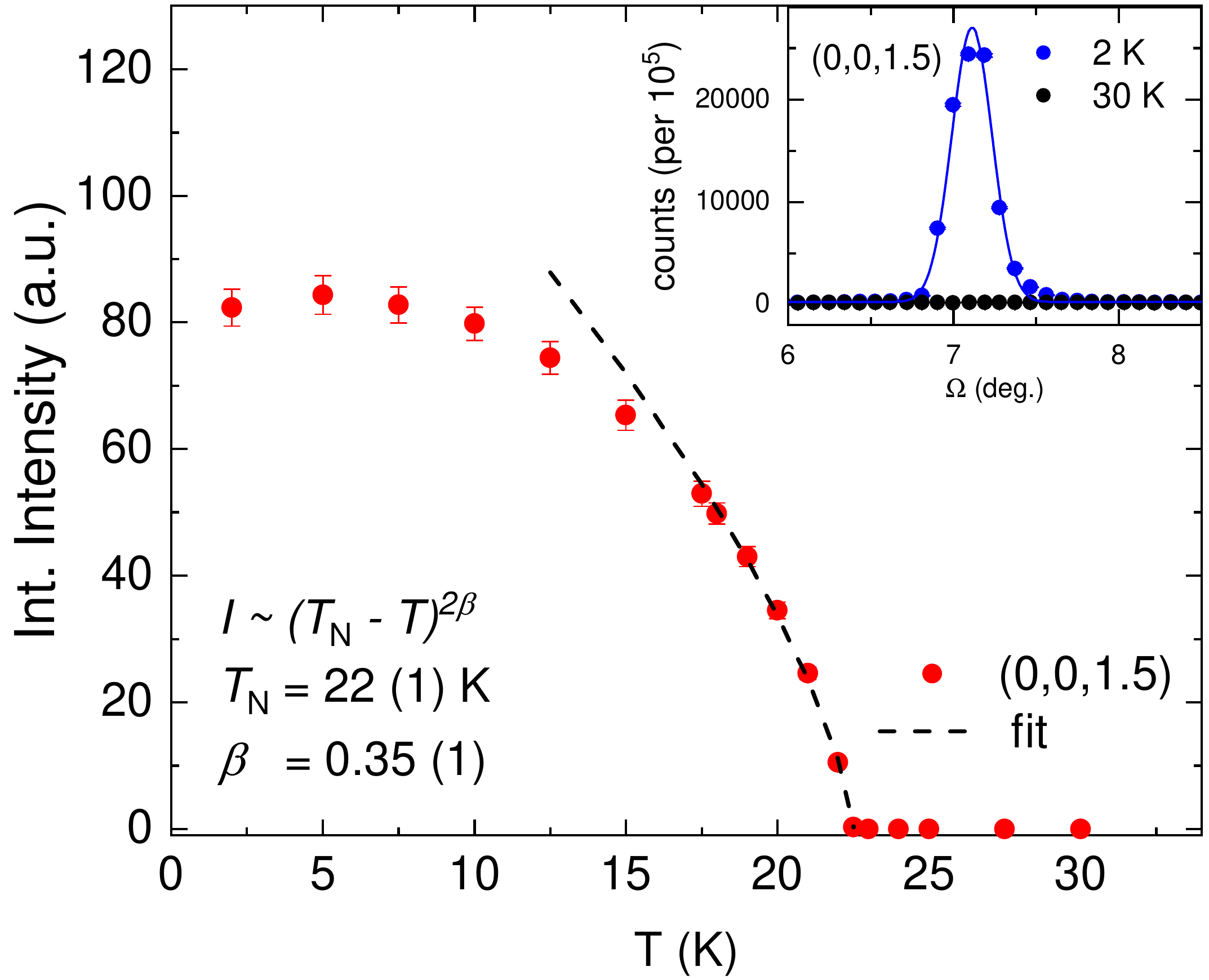}
\caption{Temperature dependence of the integrated intensity of the (0,0,1.5) magnetic Bragg peak. The dashed black curve is a fit to the data with the power law $I\sim (T_{\rm N} - T)^{2\beta}$. The inset shows the $\Omega$ scan through the magnetic (0,0,1.5) peak at 2~K and 30~K respectively. The solid blue line is the Gaussian fit to the peak at 2~K. See the text for more details.} \label{neutron_critfit}
\end{figure}


All the finite intensity magnetic peaks are observed at the general position $(H,K,L)$ + (0,0,1.5) with $H,K,L$ satisfying the reflection conditions of the $R$-3 space group and hence confirming \textbf{k} = (0,0,1.5). A few of the observed high-intensity magnetic peaks are listed in Table~\ref{tab_mag2k}. The largest diffraction intensity occurs for the magnetic Bragg peak (0,0,1.5) indicating that the Ni$^{2+}$-moments lie in the $ab$ plane which had been suggested by previous magnetization measurements~\cite{dey2020}. The temperature dependence of the integrated intensity of the commensurate reflection (0,0,1.5) in Fig.~\ref{neutron_critfit} shows finite intensity below the magnetic ordering temperature. A power law fit in the critical region using $I \propto M^2 \propto \tau^{2\beta}$ where $M$ is the order parameter and $\tau = 1 - T/T_{\rm N}$ results in \tn\ = 22(1)~K and $\beta = 0.35(1)$. The obtained value \tn\ from the power law fit agrees to the one from previous macroscopic studies~\cite{Stickler, Watanabe, Harada, dey2020}. The obtained critical parameter indicates that Ni$^{2+}$-spins in \nto\ are of 3D Heisenberg nature. 

\begin{table}[h]
  	\centering
	\caption{Observed intensities($I_{\rm obs}$) of several high-intensity magnetic peaks as measured in D10 at 2~K and their corresponding calculated intensities($I_{\rm cal}$) as discussed in the text.}\vspace{1mm}
	\begin{tabularx}{0.8\linewidth} {  >{\raggedright\arraybackslash}X  >{\centering\arraybackslash}X  >{\raggedleft\arraybackslash}X  }
		\hline \hline
		Q       &    I$_{obs}$    &   I$_{cal}$   \\ 
		\hline 
	   (0,2,2.5)      &    975(17)    &  917  \\
	   (0,2,5.5)      &    1410(27)   &  1307  \\
	   (0,-1,5.5)     &    2481(36)   &  2685  \\
	   (0,0,4.5)      &    2809(22)   &  3348  \\
	   (1,-2,-1.5)    &    1923(20)   &  2045   \\
	   (1,-2,4.5)	  &    1755(22)   &  1521   \\
	   (0,-1,2.5)     &    1787(17)   &  1965   \\
	   (-1,2,4.5)     &    1812(48)   &  1521  \\
	   (0,-1,8.5)     &    1729(109)  &  1679  \\
	   (0,0,1.5)      &    4366(21)   &  3942   \\
	  
	    \hline \hline
	\end{tabularx}
	\label{tab_mag2k}
\end{table}


\begin{figure}[ht]
\centering
\includegraphics [width=1\columnwidth] {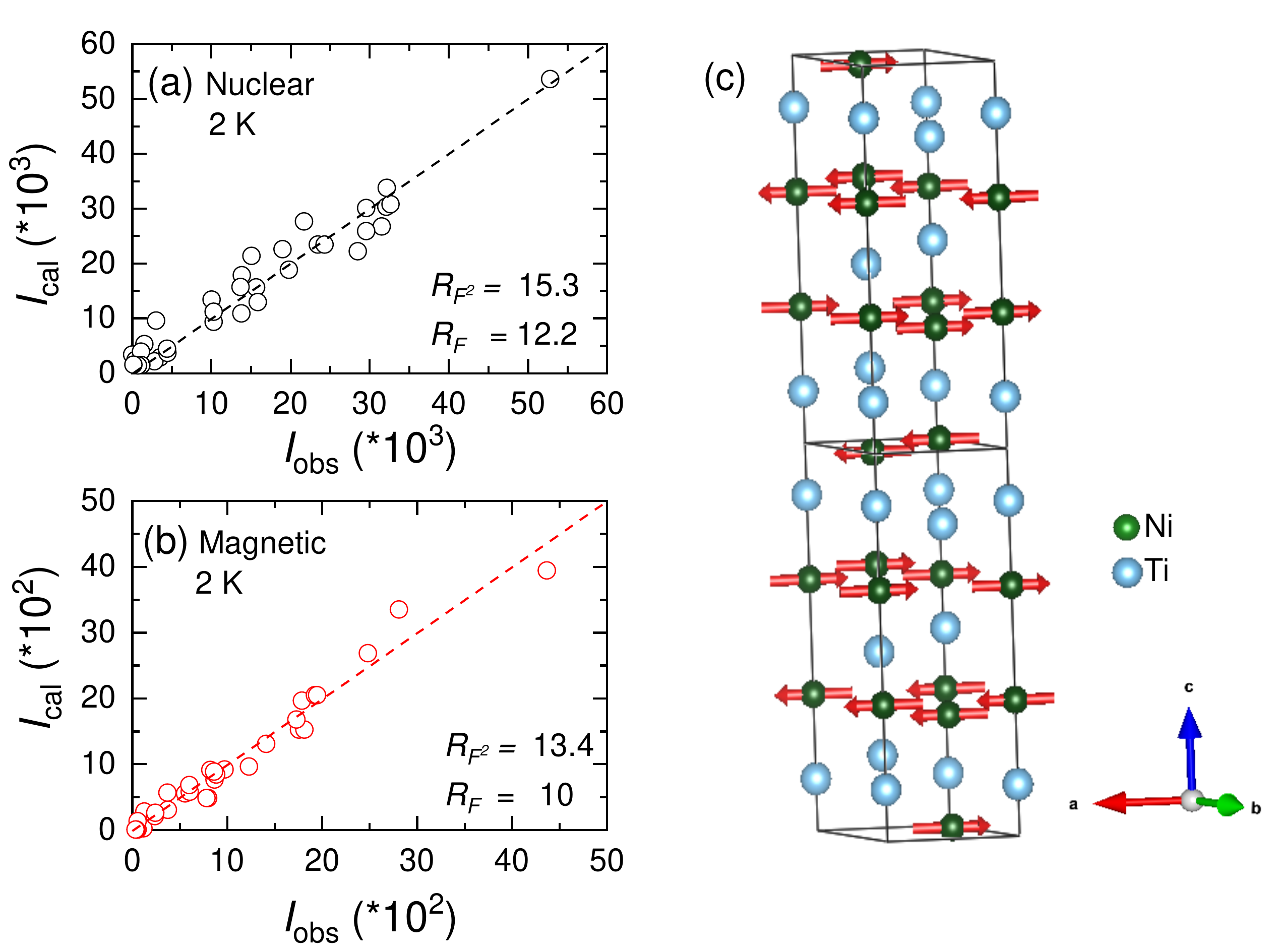}
\caption{Comparison between the observed and calculated integrated intensities of the non-equivalent nuclear (a) and magnetic (b) reflections, respectively, at 2~K, and (c) easy-plane type magnetic structure of \nto\ as determined from the refinements at 2~K.} \label{neutrostruct}
\end{figure}


The knowledge of the propagation vector \textbf{k} = (0,0,1.5) with the Ni$^{2+}$-moments lying in the hexagonal $ab$-plane points towards two possible magnetic models for \nto: (a) FM layers stacked antiferromagnetically along the $c$-axis or (b) AFM layers with the spins aligned ferromagnetically along the $c$-axis. Previous static magnetic susceptibility $\chi = M/H$ vs. $T$ measurements reveal the decrease of $\chi_{ab}$ below \tn\ whereas $\chi_c$ stays nearly constant \cite{Watanabe,dey2020}. Moreover, the magnetic model (b) implies a zero magnetic structure factor at the position $Q = (0,0,1.5)$ contrary to our observation. Hence model (a) is most suitable to describe the magnetic structure of \nto. Hence, the obtained magnetic peaks at 2~K were refined against model (a) and a very good fit was obtained as shown in Fig.~\ref{neutron_m}(b). The obtained magnetic structure of \nto\ re-confirms the structure proposed by Shirane \etal\ based on powder neutron data as early as 1959~\cite{Shirane_1959}. The calculated intensities of several high-intensity peaks are listed in Table~\ref{tab_mag2k} and the complete magnetic structure of \nto\ is schematically represented in Fig.~\ref{neutrostruct}(c). At $T=2$~K, the ordered moment amounts to 1.46(1) $\mu_\textrm{B}$. 

The crystal symmetry of the basal hexagonal planes is marked by the presence of two sets of three two-fold axes. Hence, the rotated by 120$^{\circ}$ in-plane spin-configurations are exactly equivalent leading to the presence of spin domains (i.e., three domains). Since the refinements are usually performed using the average of the integrated intensities of the equivalent reflections, the directions of the spins cannot be uniquely determined using single-crystal neutron diffraction alone, similar to the problem existing in the powder diffraction experiments~\cite{Shirane_1959}. However, excellent agreement of the integrated intensities between the equivalent reflections ($R_{int} = 1.86 \%$ indicates that there are likely three spin-domains of equal population with spins rotated by 120$^{\circ}$ in between the neighbouring domains.

\subsection{Magneto-structural-dielectric coupling}


\begin{figure}[ht]
\centering
\includegraphics [width=1\columnwidth] {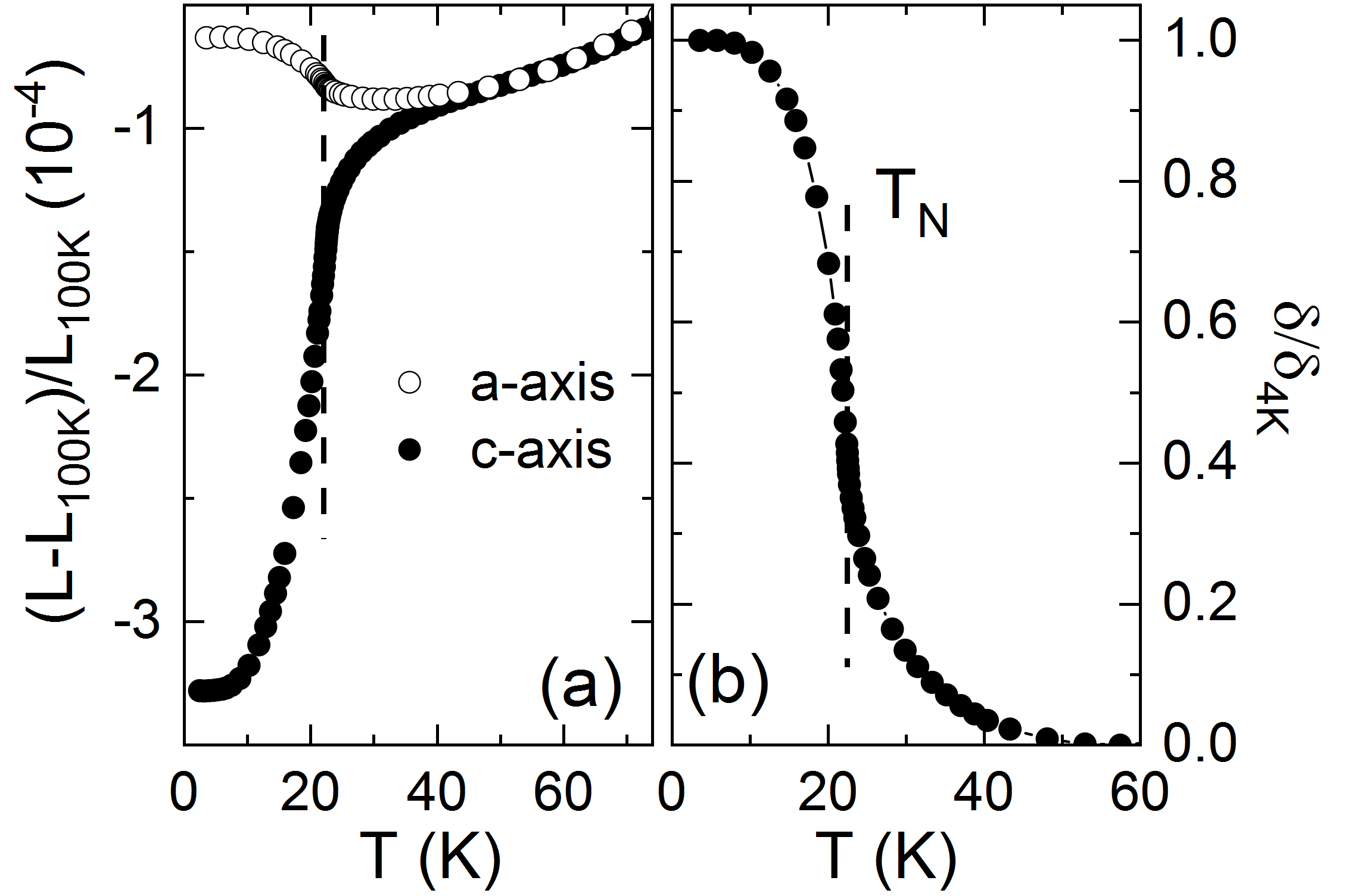}
\caption{(a) Relative length changes $dL_i^*=(L_{i}-L_i^{100K})/L_{i}^{100K}$ measured along the principle crystallographic $a$- and $c$-axis, respectively, by means of high-resolution dilatometry. (b) Normalised distortion parameter $\delta /\delta_{\rm 4K}$, with $\delta = (dL_a^*-dL_c^*)/(dL_a^*+dL_c^*)$. (c) Scaling of non-phononic linear thermal volume expansion volume ($dV'/V$) with the normalized dielectric permittivity digitzed from ref~\cite{Harada}. The vertical dashed lines indicate \tn .} \label{distortion}
\end{figure}


The magneto-structural coupling in \nto\ has been studied by means of high-resolution capacitance dilatometry. The uniaxial relative length changes $dL_i^*=(L_{i}-L_i^{100K})/L_{i}^{100K}$ ($i = a, c$) (Fig.~\ref{distortion}(a)) versus temperature show abrupt changes at \tn , i.e., shrinking of the $c$-axis and expansion along the $a$-axis, which demonstrates significant magnetoelastic coupling in \nto. At higher temperatures $T \geq 50$~K, isotropic thermal expansion coefficients results in similar rate of increase of $dL_i^*$ along the $a$- and the $c$-axis, respectively. To further elucidate lattice changes at \tn, the normalized distortion parameter $\delta/\delta_{4K}$, with $\delta = (dL_a^*-dL_c^*)/(dL_a^*+dL_c^*)$, is shown in Fig.~\ref{distortion}(b).

As evidenced by the distortion parameter, different behaviour of the $a$- and $c$-axis starts to evolve gradually below 50~K while $\delta$ sharply jumps at \tn\ (Fig. \ref{distortion}(a)). Evidently, onset of long-range AFM order is associated with a large spontaneous magnetostricton effect and it implies strong magneto-structural coupling. Large spontaneous magnetostriction has also been observed in other ilmenites such as \fto , The latter, however, shows a reversed magnetostrictive effect, i.e., there is an expansion of the $c$-axis and shrinking of the $a$-axis~\cite{charilou}. We attribute this difference to the differing magnetic ground states in \fto\ and \nto\ and corresponding variation in magneto-crystalline anisotropy. Finite distortion $\delta$ up to 50~K evidences a precursor phase with short-range order well above \tn . Due to the observed strong magnetoelastic coupling we conclude the presence of short-ranged spin correlations persisting up to twice the transition temperature. This is corroborated by previous specific heat measurements~\cite{dey2020} on \nto\ which reveal that nearly 20\% of magnetic entropy is consumed between \tn\ and 50~K. In addition, it has been shown that \textbf{q}-dependent spin-spin correlations couple to the dielectric response via the coupling of magnetic fluctuations to optical phonons, thereby causing a significant magnetocapacitive effect~\cite{Lawes_2003}. Accordingly, we conclude the significant magnetocapacitance of 0.01\% and finite magnetostriction recently observed in \nto\ well above \tn\ is due to persisting spin-spin correlations~\cite{Harada,dey2020}. 

\subsection{Spin-reorientation}


\begin{figure}[h]
\centering
\includegraphics [width=0.95\columnwidth] {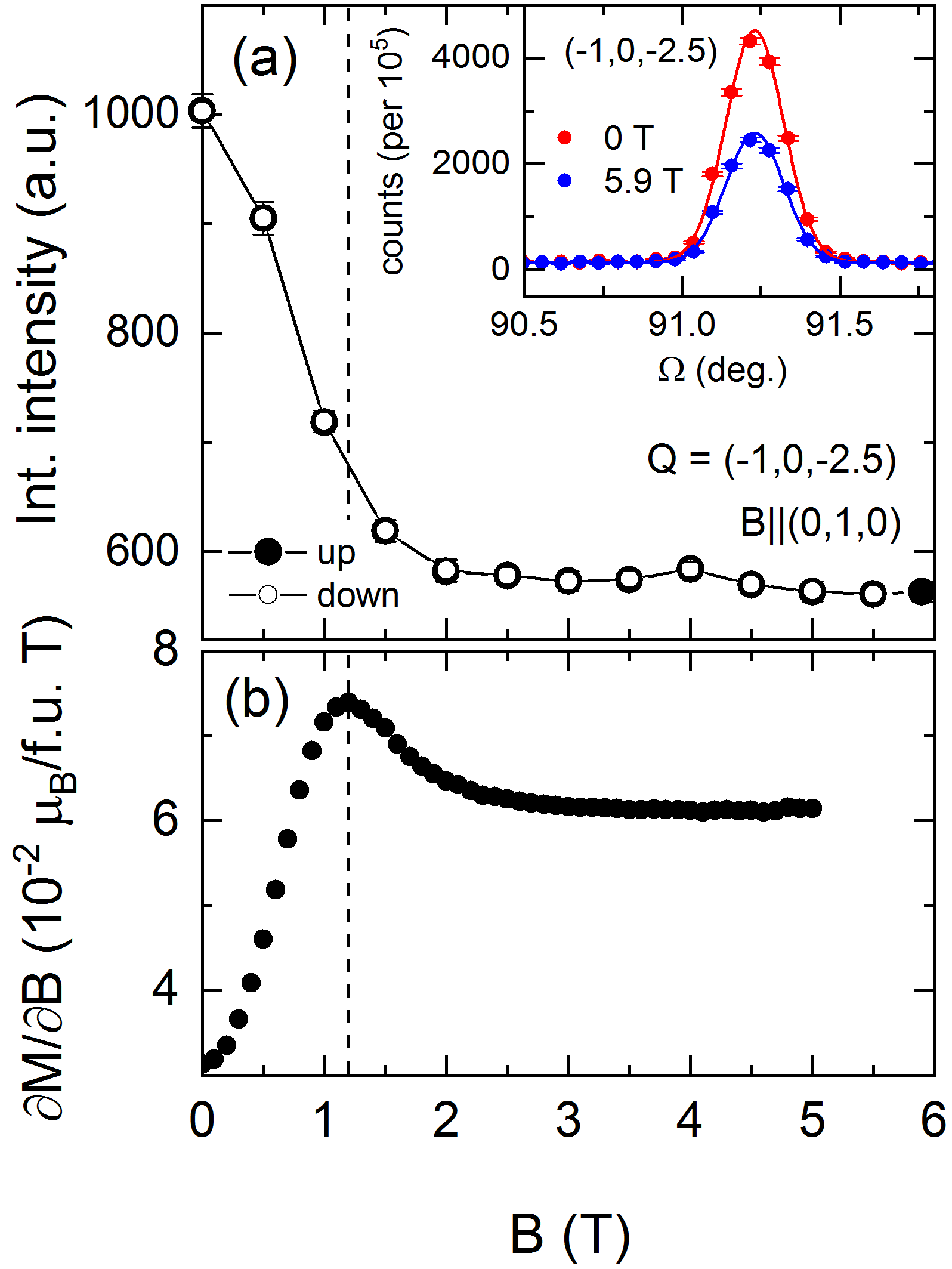}
\caption{(a) Integrated intensity of the magnetic (-1,0,-2.5) peak as a function of magnetic field (up and down) and (b) the derivative of static magnetization with respect magnetic field $\partial M/\partial B$ as a function of magnetic field (from ref. \onlinecite{dey2020}) at 2~K. The inset to (a) shows the $\Omega$-scans through the magnetic (-1,0,-2.5) peak at 0~T and 5.9~T. The solid lines in blue and red are Gaussian fits to the peaks at 0~T and 5.9~T respectively.} \label{neutron_m}
\end{figure}


The effect of magnetic fields applied within the $ab$-plane on the crystal and magnetic structure of \nto\ is studied by means of in-field neutron diffraction at 2~K. Specifically, the magnetic field is applied vertically along $b$-axis and the scattering vector lies in the $(H,0,L)$ plane. Several nuclear and magnetic reflections were measured with rocking curve scans in magnetic-fields up to 6~T. As will be discussed below, there is a considerable decrease in intensity upon application of the magnetic field for all magnetic peaks while in contrast there is no magnetic field effect on the nuclear peak intensities. A representative scan through the magnetic peak $Q$ = (-1,0,-2.5) is shown in the inset to Fig.~\ref{neutron_m}.

The magnetization curve displays a non-linear dependence on magnetic fields applied along the $ab$-plane as evidenced by the magnetic susceptibility $\chi = \partial M/\partial B$ in Fig.~\ref{neutron_m}(b). The maximum in $\chi$ at $B=1.2$~T is indicative of a spin-reorientation transition. Correspondingly, the integrated magnetic intensity (Fig. \ref{neutron_m}(a)) shows a continuous decrease in magnetic fields up to 2~T above which it stays nearly constant at a finite value. Since the magnetic neutron diffraction intensity is proportional to the component of the magnetic moments perpendicular to the scattering vector, this observation indicates that in magnetic field the spins are rotated smoothly from three magnetic domains to a single domain state with spins are aligned perpendicular to fields above 2~T. Between 2 and 6~T, negligible field dependence indicates a very small canting of spins towards magnetic field. The FWHM calculated using Gaussian fits to nuclear peaks show negligible broadening up to 6~T indicating that the magnetostriction effects on lattice parameters corresponding to the spin-reorientation is below the experimental resolution.

\subsection{Magnetostriction}


\begin{figure}[h]
\centering
\includegraphics [width=1\columnwidth] {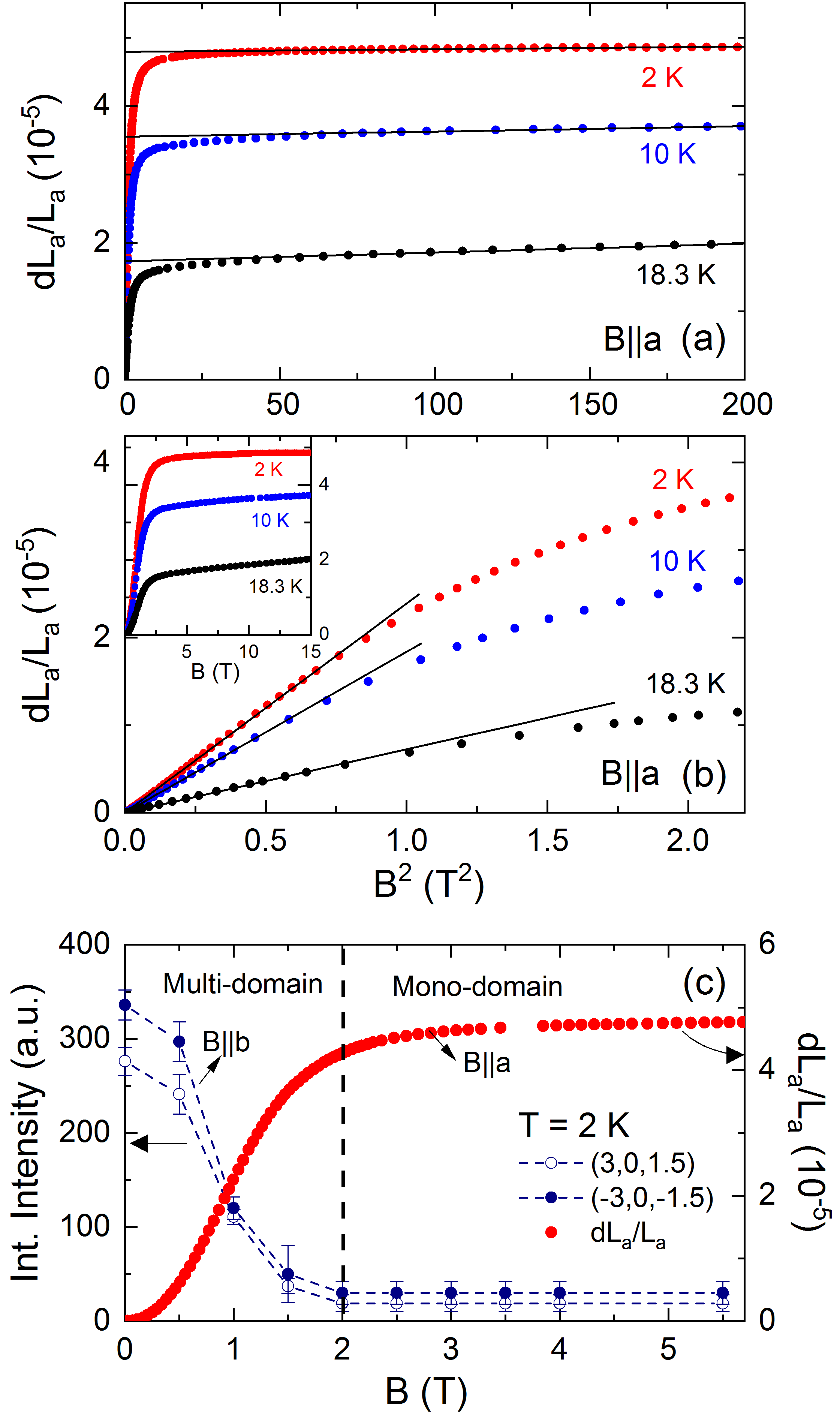}
\caption{Relative length changes $dL_{a}/L_{a}$, at different temperatures, versus the square of the magnetic field applied along the crystallographic $a$-axis for (a) magnetic fields up to 14~T, i.e., including the high-field single-domain (homogeneous) phase, and (b) for $B\leq 1.5$~T which is the low-field multi-domain phase (see the text). The solid black lines are corresponding linear fits. The inset to (b) shows the relative length changes versus applied magnetic field. (c) Integrated intensity of the equivalent magnetic Bragg peaks (3,0,1.5) and (-3,0,-1.5) vs. magnetic field applied along the $b$-axis and $(dL_{a}/L_{a})$ for fields along $a$-axis, at $T=2$~K. The vertical dashed line separates the multi-domain and the mono-domain (homogeneous) regions. See text for more details.} \label{dllvsh2}
\end{figure}


Applying magnetic fields along the $ab$-plane yields a pronounced increase of the associated lattice parameter in the low-field region ($B<B^{*} = 2$~T) while there is only small magnetostriction at higher fields (see Fig.~\ref{dllvsh2}). Magnetostriction is also reportedly small for fields applied along $c$-axis\cite{dey2020}. We conclude that this behaviour is associated with the field-driven collective rotation of spins as discussed above and evidenced by Fig.~\ref{neutron_m}. However, as will be discussed below, the magnetisation changes do not scale with magnetostriction and the maxima in $\partial M/\partial B$ and $\partial L_a/\partial B$ do not match each other (see Fig.~\ref{scaling}(a)). The magnetostriction data hence do not correspond to what is expected for a thermodynamic spin-reorientation transition. Instead, the presence of domains has to be involved and in the following we will present clear evidence that the data represents the change from a low-field multi-domain state to a high-field uniform mono-domain one. 

In order to further investigate the effect of in-plane magnetic fields, the intensity evolution of two equivalent magnetic Bragg peaks (3,0,1.5) and (-3,0,-1.5) belonging to two different magnetic domains is displayed in Fig.~\ref{dllvsh2}(c). In the multidomain state, the antiferromagnetic vector is uniform within a single domain and has different directions in different domains. The observed isotropic decrease in intensity of both magnetic peaks upon application of the magnetic field implies that the spins of both domains rotate perpendicularly to the external field direction. The spin-rotation process is completed at $B^*$ which hence signals the formation of a spin-reoriented monodomain state. Accordingly, no significant changes in the peak intensities are observed above $B^*$ up to 6~T. 

\section{Discussion}

\begin{figure}[h]
\centering
\includegraphics [width=1\columnwidth] {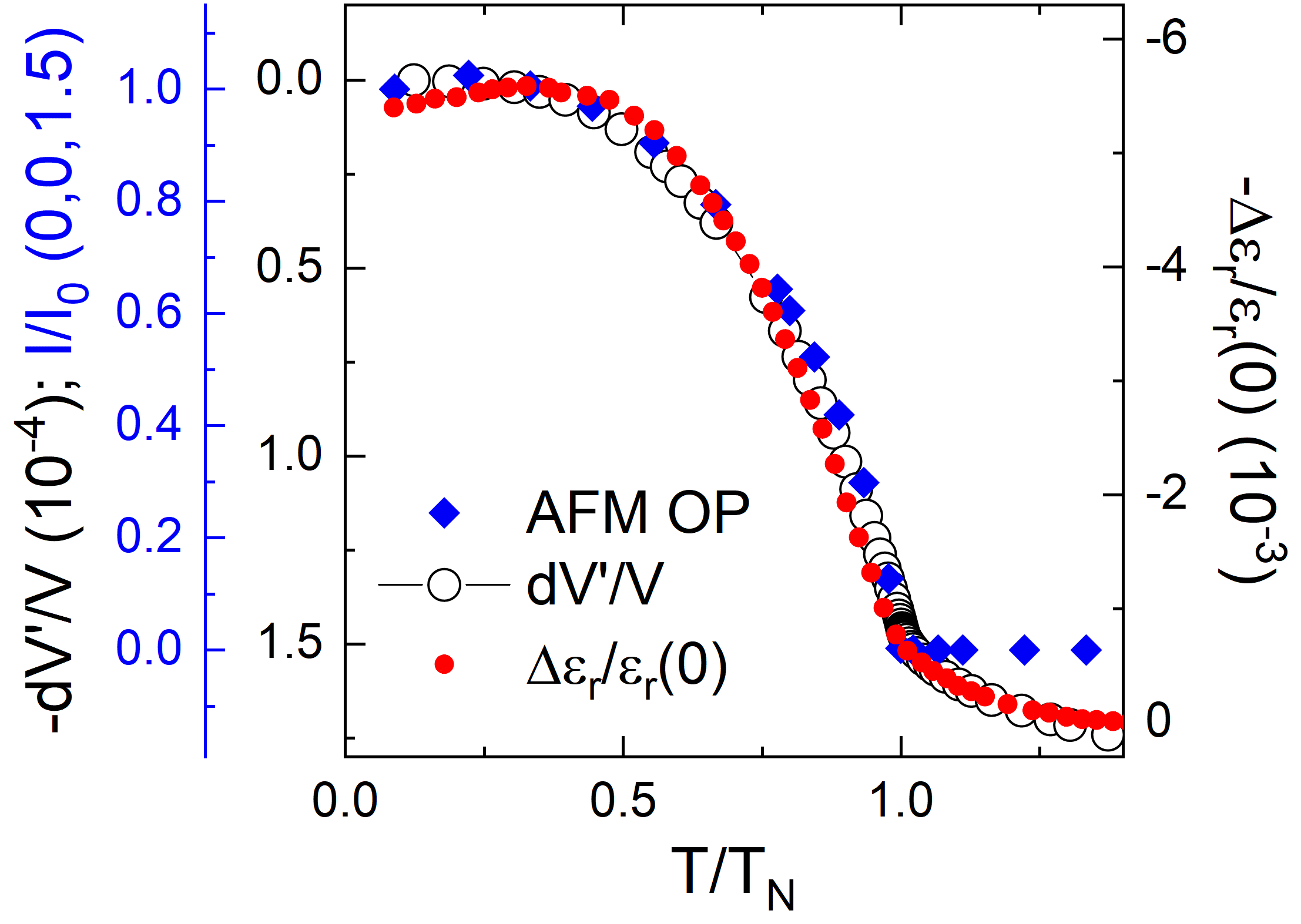}
\caption{Temperature dependence of the square of AFM order parameter, i.e., the normalised integrated intensity of the (0,0,1.5) magnetic Bragg peak, the negative non-phononic volume changes $dV'/V$, and the normalized dielectric permittivity digitized from Ref~\onlinecite{Harada}.} \label{scale}
\end{figure}

Comparison of the magnetic order parameter and the relative volume changes with the reported data of the dielectric function by Harada \etal~\cite{Harada} elucidates the coupling mechanism between the lattice and the dielectric degrees of freedom in \nto . As displayed in Fig.~\ref{scale}(a), the non-phononic relative volume changes $dV'/V = 2(dL_a/L_a) + (dL_c/L_c)$  which are obtained by subtracting the phononic contribution from $dV/V$ (cf. Ref.~\onlinecite{dey2020}) show a very similar temperature dependence, below \tn , as the normalized dielectric permitivitty. Note, that the polycrystalline sample studied in Ref.~\onlinecite{Harada} displays a slightly lower \tn\ than the single crystals studied at hand. Note, that in general length changes can directly affect the experimentally measured permitivitty via the relation $\epsilon = Cd/\epsilon_0 A$, where $C,\epsilon_0, d$ and $A$ are sample capacitance, vacuum permittivity, sample thickness and area, respectively. However, the changes in sample dimensions at \tn\ are on the order of 10$^{-4}$, while the relative change in permitivitty an order higher, implying that spontaneous magnetostriction is not the driving mechanism for the observed dielectric changes at \tn. Interestingly, the normalized dielectric permitivitty varies as square of the antiferromagnetic order parameter '$L$' represented by the normalized integrated intensity of the magnetic (0,0,1.5) Bragg peak in Fig.~\ref{scale}(a). In order to discuss this, we recall the Landau expansion of the free energy $F$, in terms of polarization $P$, and the sub-lattice magnetization $L$ at zero magnetic-field \cite{sparks2014}:

\begin{equation}
F = F_0 + \alpha P^2 + \alpha'L^2 + \beta PL + \gamma P^2L^2 - PE  
\end{equation}

The dielectric function is obtained as $\partial^{2}F/\partial P^2 = \epsilon \propto \gamma L^2$. Hence Fig.~\ref{scale}(a) qualitatively evidences the presence of magnetodielectric coupling in \nto. On top of the spin and dielectric changes, the structural changes exhibits similar behaviour below \tn. Previously reported magnetic Gr{\"u}neisen analysis\cite{dey2020} evidences that the entropic changes at \tn\ to be of purely magnetic nature. In our opinion the spin-phonon coupling is responsible for observed dielectric changes at \tn. In the presence of spin-phonon coupling the phonon frequency $\omega$ can be affected by spin-correlation as $\omega = \omega_0 + \lambda <S_i.S_j>$ resulting in modification of permitivitty via the Lyddane-Sachs-Teller equation $\epsilon_0 = \omega{_L}{^2}/\omega{_T}{^2}\epsilon_{\infty}$, where $\epsilon_0$ and $\epsilon_{\infty}$ are the permitivitty at zero frequency and optical frequency, respectively and $\omega{_L}{^2}$ and $\omega{_T}{^2}$ are the long-wavelength longitudinal and transverse optical phonon modes respectively.

It is noteworthy that apart from spontaneous magnetostriction, an exchange-striction (ES) mechanism may in principle also lead to spontaneous lattice deformation at \tn\ and be a potential source for dielectric anomaly at \tn. Magnetodielectricity fueled by ES mechanism have been observed in several systems for example Y$_2$Cu$_2$O$_5$\cite{Adem_2009} and TeCuO$_3$\cite{Lawes_2003}. In \fto\, a combination of ES and magnetostriction mechanisms have been suggested for the spontaneous lattice deformation at \tn\cite{charilou}. In particular for \nto\ an ES mechanism would imply a change in Ni-O-Ni bond angle in the $ab$ plane closer to 90$^\circ$ favouring ferromagnetic super-exchange. However, diffraction experiments reveal that the bond angle increases from 90.34$^\circ$ at 100~K to 90.36$^\circ$ at 2~K (supplementary Fig. 2), contrary to predictions of ES. Hence, ES mechanism is excluded as the origin of lattice distortion at \tn in \nto.

The crystallographic symmetry of the easy hexagonal plane in \nto\ suggests the presence of three domains with spins rotated by 120\degree\ in different domains. Such a spin structure with three domains is often observed in easy-plane-type hexagonal antiferromagnets such as CoCl$_2$, NiCl$_2$, and BaNi$_2$V$_2$O$_8$~\cite{Wilkinson_1959,Knafo}. In \nto , the magnetostriction data imply that the field-driven changes of the domain structure is associated with structural changes. Indeed, orientational AFM domains are magnetoelastic in nature~\cite{Gomonay_1999,Gomonay_2002} and have previously been observed in various systems, for example in cubic antiferromagnets RbMnF$_3$~\cite{Shapira_1978}, KNiF$_3$ and KCoF$_3$~\cite{Tanner_1978,Tanner_1979}, NiO~\cite{Yamada_1966}, iron-group dihalides CoCl$_2$~\cite{Kalita_2000} and NiCl$_2$~\cite{Kalita_2002}, the quasi-two-dimensional AFM BaNi$_2$V$_2$O$_8$~\cite{Knafo}, YBa$_2$Cu$_3$O$_{6.3}$~\cite{Gomonay_2002} etc. In particular, Kalita and co-workers have developed phenomenological theories describing the effect of domain re-distribution on the magnetostriction for CoCl$_2$ and NiCl$_2$\cite{Kalita_2000,Kalita_2001,Kalita_2002,Kalita_2004,Kalita_2005}. Note, that both NiCl$_2$ and CoCl$_2$ are easy-plane-type antiferromagnets with similar crystalline symmetry, i.e., trigonally distorted octahedral environment surrounding metal ions, similar to \nto\ and \cto~\cite{Wilkinson_1959,Lines_1963}. In the following, we will describe the field-dependency of the lengths changes in \nto\ based on the phenomenological theories developed by Kalita and co-workers. 

Both at low magnetic fields $B||a \leq 1$~T and at high fields the field-induced striction $dL_{a}/L_{a}$ varies as the square of the applied magnetic field as shown in Fig. \ref{dllvsh2}(a,b). In the latter, i.e., the mono-domain state, this is predicted by calculating the equilibrium elastic strain by energy minimization of the magnetoelastic and the elastic contributions to the free energy~\cite{Kalita_2000,Kalita_2002}. The magnetostriction in the mono-domain state is described by 

\begin{equation}
(dL_{a}/L_{a})(T,B) =  \alpha (T) (B)^2 + (dL_{a}/L_{a})_s(T,B=0~{\rm T}) \label{eqmono}
\end{equation}

where $\alpha (T)$ is the temperature dependent slope and $(dL_{a}/L_{a})_s(T,H=0)$ is the spontaneous magnetostriction of the mono-domain state obtained by extrapolating the linear fit to $B=0$~T. Eq.~\ref{eqmono} fits well with $dL_{a}/L_{a}$ at different temperatures as shown by the solid black lines in Fig.~\ref{dllvsh2}(a). The obtained fit parameters are listed in Table ~\ref{table_dll_fit}. $(dL_{a}/L_{a})_s$ corresponds to a hypothetical spontaneous striction that would be observed if the magnetoelastic domains did not appear at low fields, i.e., if the total spontaneous magnetostriction was not compensated on the whole by summation of strains in different directions in each of the domains. 

The magnetostrictive response upon application of magnetic fields in the multi-domain state is governed by domain-wall motion. Specifically, magnetostriction is large due to the associated facilitated rotation of spins. The motion of magnetoelastic domain-walls is predominantly reversible in nature~\cite{Tanner_1978,Gomonay_1999} and the associated lengths changes again exhibit a square-dependence on the magnetic field which is expressed by

\begin{equation}
(dL_{a}/L_{a})(T,H) =  (dL_{a}/L_{a})_s(T,H=0) (H/H_{\rm d})^2. \label{eqmulti}
\end{equation}

Here, $H_{\rm d}$ is an empirical parameter obtained from the fits (see Table~\ref{table_dll_fit}). As shown in Fig.~\ref{dllvsh2}(b), the experimental data are well described by Eq.~\ref{eqmulti} which is in-line with the predictions of phenomenological models~\cite{Tanner_1978,Kalita_2002}. Although the magnetoelastic domains are predominately reversible in nature, a small irreversibility may arise due to pinning of domain walls by crystal defects and imperfections in the crystals. A small remanent striction amounting to $\sim 1.6\times10^{-6}$, at $T=2$~K, is indeed observed in our data (see the Supplement Fig. 1) which indicates the presence of predominately mobile domain walls~\cite{Gomonay_1999} in \nto. 

\begin{table}[h]
	\centering
	\caption{Parameters obtained from fits to the magnetostriction data (Fig.~\ref{dllvsh2}(a,b)) using Eqs.~\ref{eqmono} and \ref{eqmulti}. $(dL_a/L_a)_s$ is the spontaneous magnetostriction (see the text).}\vspace{1mm}
\begin{tabularx}{1\linewidth} {>{\raggedright\arraybackslash}X  >{\centering\arraybackslash}X  >{\centering\arraybackslash}X  >{\centering\arraybackslash}X }
	  \hline \hline 
		$T$  & $(dL_a/L_a)_s (10^{-5})$ & $H_{\rm d}$~(T) & $\alpha$ ($10^{-9})$~(T$^2$) \\ 
		\hline
	    2~K    &  4.79   &      1.41         &  3.8     \\
		10~K   &  3.55   &      1.38         &  7.6	       \\
	   18.3~K  &  1.73   &      1.55         &  12.8  \\
      \hline \hline
	\end{tabularx}
	\label{table_dll_fit}
\end{table}


\begin{figure}[h]
\centering
\includegraphics [width=1\columnwidth] {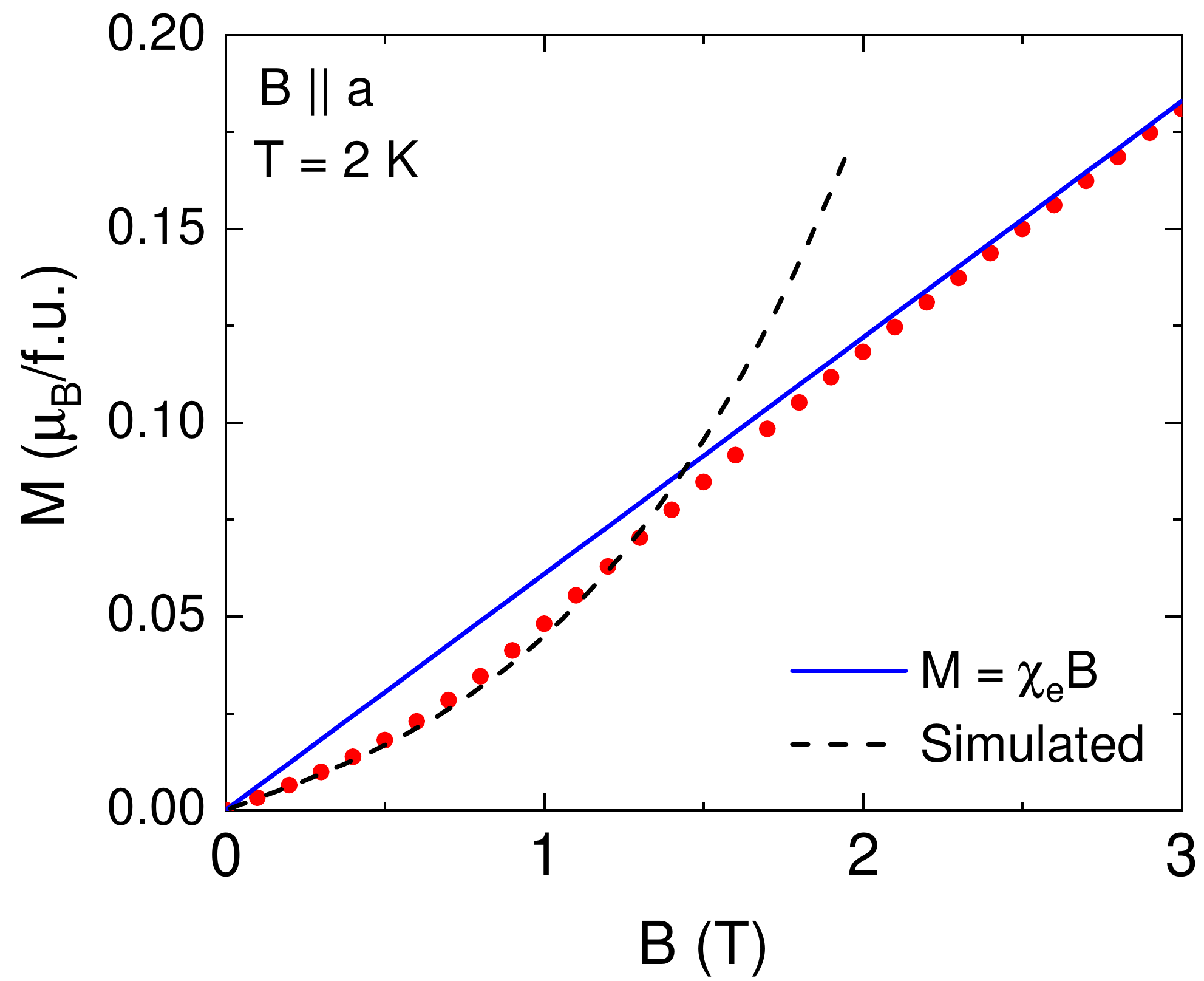}
\caption{Magnetization $M$, at $T=2$~K, versus applied magnetic field $B||a$-axis. The solid blue line represents a linear fit to $M$ in the high-field region and the dashed black line shows a simulation to $M$ at low fields (see the text for more details). } \label{simM}
\end{figure}


Unlike uniaxial antiferromagnets which show an abrupt magnetization jump at the spin-flop transition as, e.g., observed in \mto~\cite{YAMAUCHI_1983}, the magnetization in \nto\ follows a sickle-shaped field dependence in the non-flopped phase and the reorientation transition is associated with smooth right bending in $M$ vs.$B$ (see Fig. \ref{simM}). Such characteristic smooth non-linear variation of magnetization in low-fields is a manifestation of the multi-domain state where spin-reorientation takes place gradually by displacement of domain walls~\cite{Tanner_1979}. This is described~\cite{Kalita_2005} by

\begin{equation}
M = (1/2)\chi_e B [1 + (B/B_{\rm d})^2]   \label{sim} 
\end{equation}

where $\chi_e$ is the high-field magnetic susceptibility. A linear fit to the $M$ vs. $B$ curve~\cite{dey2020} at $B>4$~T yields $\chi_e = 0.06$~\mb/f.u.T which is represented by the solid blue line in Fig.~\ref{simM}. Using $H_{\rm d}$ from the analysis of the magnetostriction data (see Table~\ref{table_dll_fit}) enables to deduce the field dependence of $M$. The simulation using Eq.~\ref{sim} is shown by the dashed line in Fig.~\ref{simM}. It yields a good description of the field-driven evolution of the magnetization in the multi-domain state, thereby further confirming the applied phenomenological model. The blue line in Fig.~\ref{simM} represents the expected magnetization in a single-domain easy-plane AFM with no in-plane anisotropy.


\begin{figure}[h]
\centering
\includegraphics [width=1\columnwidth] {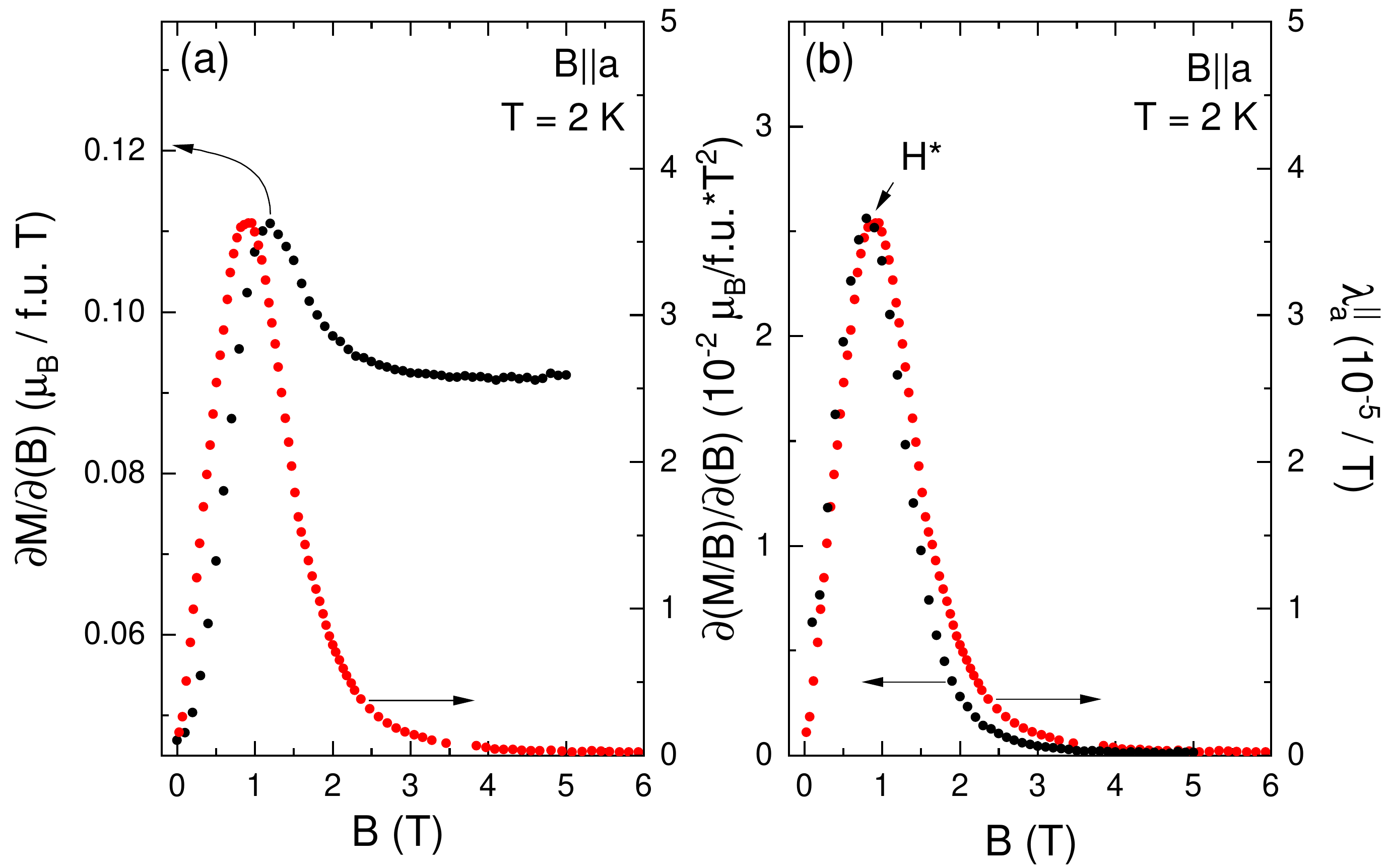}
\caption{ (a) Scaling of $\partial M$ / $\partial B$ , (b) $\partial (M/B)$ / $\partial B$  and $\lambda_{a}^{\parallel}$ versus $B$ at $T=2$~K.}\label{scaling}
\end{figure}


The field-driven disappearance of the multi-domain state yields different behaviour of the magnetic susceptibility $\partial M/\partial B$ and the magnetostriction $\partial L/\partial B$. This is demonstrated in Fig.~\ref{scaling}(a) where the derivative of the magnetization and the longitudinal magnetostriction coefficient $\lambda_{a}^{||} = (1/L_a)\partial L_a/ \partial (\mu_0H)$ are shown at $T=2$~K as a function of $B$. The data are scaled to match the corresponding peak values. 
According to the Ehrenfest relation

\begin{equation}
\partial B^*/ \partial p_i = V_m \Delta\lambda_i/\Delta[\partial M/\partial (\mu_0H)]
\end{equation}

Using molar volume $V_m =$ 42.01~cm$^3$/mol and $B^*=$ 0.8~T (Fig.~\ref{scaling}(b)), we obtain the normalized pressure dependency $(1/B^*) \partial B^*/ \partial p = 0.8$~kbar$^{-1}$. Positive magnetostriction in the mono-domain phase reveals that (see also Fig.~\ref{dllvsh2} (a)) for each domain the in-plane distortion in magnetic field is such that the lattice expands perpendicular to the spin-direction. Hence, applying a uniaxial pressure $p$ will induce an anisotropy in plane favouring domains with spins nearly parallel to $p$ in the multi-domain phase.

 The scaling of $\partial (M/H) /\partial (\mu_0H)$ and $\lambda_{a}^{||}$ at 2~K in Fig.~\ref{scaling}(b) shows that the quantities vary proportional to each other in the multi-domain state peaking at $B^*$. The proportional variation  $d(m/H)/dH \sim \lambda_{a}^{||}$ is consistent with equation 3 and is manifestation of magnetoelastic nature of the domains. The behaviour is expected from phenomenological theories of magnetoelastic domains which describe the variation of magnetization and length changes by means of a single domain co-alignment parameter and it's variation with magnetic field \cite{Kalita_2004}. 

Apart for large magneto-crystalline anisotropy which dictates the easy-plane spin structure in \nto, an additional small in-plane anisotropy may arise due to frozen strains in the domain walls\cite{Lozenko_1974,Weber_2003}. Small in-plane anisotropy has been previously observed in other easy-plane type antiferromagnets like the dihalides NiCl2 ($\sim$ 0.3~T) and CoCl2 ($\sim$ 0.8~T) by means of low-frequency resonance experiments\cite{Lozenko_1974} and in \cto($\sim$ 1 meV) by means of INS experiments \cite{elliot_2020} respectively. Although bond anisotropic exchange interaction pinning the order parameters to the crystal axes \cite{Yuan_2020} was suggested as the responsible mechanism for small in-plane gap in \cto\ we believe that a small in-plane anisotropy to be present in \nto\ corresponding to magnetoelastic domain walls.

\section{Summary}

In summary, we have studied in detail the magneto-structural coupling in magnetodielectric \nto\ by means of single crystal neutron-diffraction and high-resolution dilatometry. Zero-field neutron diffraction reveals multidomain A-type spin antiferromagnetic ordering with preservation of crystallographic $R$-3 symmetry down to 2~K. Zero-field thermal expansion measurements reveals spontaneous lattice deformation at \tn. The dielectric permitivitty $\epsilon$ scales with the square of magnetic order parameter $L$ in line with predictions of Landau theory and hence indicating finite magnetodielectric coupling in \nto. Our analysis suggests the presence of spin-phonon coupling as a responsible mechanism for dielectric anomaly at \tn\ in \nto. In-field neutron diffraction shows the evolution of  magnetic domains with spins perpendicular to the applied field. The effect of magnetic domains on magnetostriction have been discussed in light of phenomenological multi-domain theories. We see magnetization and magnetostriction scale with each other in the multidomain state revealing strong coupling of spins to the lattice.    

\begin{acknowledgements}
This work has been performed in the frame of the International Max-Planck School IMPRS-QD. We  acknowledge support by BMBF via the project SpinFun (13XP5088) and by Deutsche Forschungsgemeinschaft (DFG) under Germany’s Excellence Strategy EXC2181/1-390900948 (the Heidelberg STRUCTURES Excellence Cluster) and through project KL 1824/13-1.  

\end{acknowledgements}


\end{document}